\documentclass[prb,twocolumn,showpacs,superscriptaddress]{revtex4}%
\usepackage{amsfonts}
\usepackage{amsmath}
\usepackage{amssymb}
\usepackage{graphicx}%
\begin{document}

\title{Adsorption and dissociation of H$_{2}$O monomer on ceria(111): Density functional theory calculations}
\author{Shuang-Xi Wang}
\affiliation{State Key Laboratory for Superlattices and
Microstructures, Institute of Semiconductors, Chinese Academy of
Sciences, P. O. Box 912, Beijing 100083, People's Republic of China}
\affiliation{Department of Physics, Tsinghua University, Beijing
100084, People's Republic of China} \affiliation{LCP, Institute of
Applied Physics and Computational Mathematics, P.O. Box 8009,
Beijing 100088, People's Republic of China}
\author{Ping Zhang}
\thanks{Corresponding author; zhang\_ping@iapcm.ac.cn}
\affiliation{LCP, Institute of Applied Physics and Computational
Mathematics, P.O. Box 8009, Beijing 100088, People's Republic of
China}
\author{Shu-Shen Li}
\affiliation{State Key Laboratory for Superlattices and
Microstructures, Institute of Semiconductors, Chinese Academy of
Sciences, P. O. Box 912, Beijing 100083, People's Republic of China}

\pacs{68.43.Bc, 68.43.Fg, 68.43.Jk, 68.47.Gh}

\begin{abstract}
The adsorption properties of isolated H$_{2}$O molecule on
stoichiometric and reduced ceria(1111) surfaces are theoretically
investigated by first-principles calculations and molecular dynamics
simulations. We find that the most stable adsorption configurations
form two hydrogen bonds between the adsorbate and substrate. The
water molecule is very inert on the stoichiometric surface unless up
to a high temperature of 600 K. For the reduced surface, we find
that the oxygen vacancy enhances the interaction. Moreover,
simulations at low temperature 100 K confirm that it is facilitated
for water to dissociate into H and OH species.
\end{abstract}

\maketitle

\section{INTRODUCTION}

Ceria (CeO$_{2}$) is a crucial material for catalytic applications
due to its remarkable properties \cite{Trovarelli1996,Fu2003}. One
of its important application is that ceria can effectively promote
the water-gas shift
reaction, $\mathrm{CO}+\mathrm{H_{2}O}\rightarrow\mathrm{CO_{2}}%
+\mathrm{H_{2}}$, by which converting harmful carbon monoxide to the less
harmful carbon dioxide, and generating hydrogen. A full account of the
involved mechanism has not been achieved, while the interactions between water
and ceria surface may plan a key role during the reaction.

Therefore, a systematic study of the fundamental interactions
involved in the water-ceria system is much desirable. The
oxygen-terminated ceria(111) surface, which is most often found in
experiments \cite{Siokou1999}, has been identified to be
energetically the most stable
\cite{Skorodumova2004,Yang2004,Jiang2005,Fronzi2009a}. The formation
of oxygen vacancies in ceria, hence reducing the cerium from
Ce$^{4+}$ to Ce$^{3+}$, is closely related to its interesting
properties and important implications, such as oxygen storage
capacity of ceria \cite{Trovarelli1996}. A heavily debated question
is on the favorable position of the oxygen vacancy. Theoretically,
different research methods including density-functional theory (DFT)
calculations, have given different favorable vacancy sites (
on-surface or on-subsurface) \cite{Esch2005,Torbrugge2007}. For the
interaction between adsorbate and ceria surface, obviously, the
on-surface oxygen vacancy is much concerned.

Experimentally, it has been found that water on reduced ceria
powders can effectively oxidize Ce$^{3+}$ into Ce$^{4+}$, with the
H$_{2}$ production \cite{Otsuka1983,padeste1993,Kundakovic2000}.
More recently, evidences have been provided by Henderson \emph{et
al.} \cite{Henderson2003} for that the water may split at low
temperature below 170 K, and exposure of water at 650 K results in
additional surface reduction of ceria instead of oxidation. In
addition, they offered three potential adsorption geometries for
water on the stoichiometric cerial(111) surface, namely, the
so-called C$_{2v}$ geometry without hydrogen bond between the water
and the substrate (no-H-band), single hydrogen bond (one-H-bond),
and two hydrogen bonds (two-H-bond) configurations, needing to be
examined by theoretical calculations.

Several theoretical works have investigated the H$_{2}$O-ceria
interaction by DFT calculations. Kumar \emph{et al.}
\cite{Kumar2006} reported that the water adsorption on the
stoichiometric ceria(111) surface favors one-H-bond configuration,
consistent with the later literature results
\cite{Chen2007,Yang2010}. While this conflicts with the results
obtained by Fronzi \emph{et al.} \cite{Fronzi2009}, who suggested a
two-H-bond structure. For water on reduced (with on-surface oxygen
vacancy) ceria(111) surface, the calculated stable adsorption
structures varied from no-H-bond \cite{Fronzi2009,Yang2010}, to
one-H-bond \cite{Kumar2006} configuration, while Chen \emph{et al.}
\cite{Chen2007} found no binding between water and reduced surface.
Moreover, Kumar \emph{et al.} \cite{Kumar2006} predicted that
oxidation of the reduced surface by water with the production of
hydrogen gas is weakly exothermic, while  Yang \emph{et al.}
\cite{Yang2010} and Fronzi \emph{et al.} \cite{Fronzi2009} argued
that hydroxyl surface forms upon water dissociating.

Such discrepancies may partially arise from the different unit cell
adopted by different authors. Actually, the water coverage adopted
in previous works ranges 0.25--1.0 ML. Too high water coverage would
be problematic because it may mask the true H$_{2}$O-ceria
interaction, by the dipole-dipole interactions between adjacent
water molecules. Therefore, a systematical study of water on
stoichiometric and reduced  ceria(111) surfaces at lower coverage,
which minimizes the lateral interaction between the adsorbates, is
indispensable to shed a light on such discrepancies for a proper
understanding of water adsorption on ceria(111) surface.

The DFT+\emph{U} calculations based on first-principles method have
proved to be an effective approach for studying the strong
correlation materials consisting \emph{d} and \emph{f} electrons,
compared with the plain DFT method \cite{Anisimov1997}. For bulk
ceria, there has been a vast literature dealing with this issue,
despite the controversial determination of the Hubbard parameter
\emph{U}
\cite{Kumar2006,Jiang2005,Fabris2005,Andersson2007,Loschen2007}. For
the surface properties of ceria, however, by declaring its
negligible influence, little work takes into consideration the
\emph{U} parameter
\cite{Yang2004,Fronzi2009a,Skorodumova2004,Jiang2005}. Nevertheless,
it has been revealed that the plain DFT method failed to correctly
describe the behavior of the water molecule on reduced ceria surface
\cite{Fronzi2009,Yang2010}. Due to the localization of \emph{f}
electrons, especially for reduced ceria surfaces with oxygen
vacancies, to describe more accurately the excess electrons, strong
correlation effect must be considered \cite{Kumar2006,Fabris2005},
hence the DFT+\emph{U} approach is required. In addition to the
static DFT calculation, \emph{ab initio} molecular dynamics (AIMD)
simulations can effectively demonstrate the dynamical properties of
a system when considering temperature, which can be compared with
the existing experimental measurements. Moreover, the use of AIMD
can be a beneficial complement to the static calculations,
especially for systems showing multiple low-lying minima in energy
\cite{Carrasco2008}.

In the present work, we systematically study the properties of
isolated water adsorption on stoichiometric and reduced ceria(111)
surfaces at low coverage by means of first-principles methods. We
perform the microscopic static DFT calculations aiming at
elucidating the electronic structure and static stability for the
water adsorption structures. Besides, AIMD simulations at various
temperatures are also conducted to investigate the dynamical
properties of these systems.

\section{Computational method}

The calculations are performed within DFT using the projector-augmented wave
(PAW) method, as implemented in the plane-Wave based Vienna \textit{ab-initio}
simulation package (VASP) \cite{Kresse1996}. The cutoff energy for the plane
wave expansion is set to 400 eV. For the bulk ceria, the exchange and
correlation effects are treated within both the local density approximation
(LDA) and the PBE \cite{Perdew1996} generalized gradient approximation (GGA).
For the ceria surface, however, only LDA is adopted. Here the cerium
5\emph{s}, 5\emph{p}, 5\emph{d}, 4\emph{f}, and 6\emph{s}, and the oxygen
2\emph{s} and 2\emph{p} electrons are treated as valence electrons. We use the
DFT+\emph{U} formalism formulated by Dudarev \textit{et al.}
\cite{Dudarev1998} to account for the strong on-site Coulomb repulsion amongst
the localized Ce 4\emph{f} electrons. In this scheme the on-site two-electron
integrals, expressed in terms of two parameters, i.e., the Hubbard parameter
\emph{U}, which reflects the strength of the on-site Coulomb interaction, and
the parameter \emph{J}, which adjusts the strength of the exchange
interaction, are combined into a single parameter $\mathrm{U}_{\mathrm{eff}%
}=U-J$, indicating that only the difference between \emph{U} and
\emph{J} is significant. For the bulk ceria, the Coulomb \emph{U} is
treated as a variable, while the exchange energy \emph{J} is set to
be 0.7 eV. Then the parameter U$_{\mathrm{eff}}$ is fixed at a
reasonable specific value for the surface calculations.

The oxygen-terminated ceria(111) surface is modeled by a slab
composing of six atomic layers and a vacuum region of 15 \AA . A (3
$\times$ 3) surface unit cell, in which each monolayer contains nine
atoms, is adopted in the study of the H$_{2}$O adsorption. The
isolated H$_{2}$O molecule is placed on one side of the slab,
equivalent to 1/9 coverage. Our test calculations have showed that
such low a coverage is sufficient to avoid the interaction between
adjacent H$_{2}$O molecules, hence makes the true H$_{2}$O-ceria
interaction emerge. During our calculations, the bottom two atomic
layers of the substrate are fixed, and other atoms as well as the
H$_{2}$O molecule are free to relax until the forces on the ions are
less than 0.03 eV/\AA . Integration over the Brillouin zone is done
using the Monkhorst-Pack scheme \cite{Monkhorst1976}. For bulk
ceria, we use a $9\times9\times9$ grid points, and for the surface
we use a $3\times3\times1$ grid points.

The calculations of the energy barriers for the water diffusion and
dissociation processes are performed using the climbing image nudged elastic
band (CI-NEB) method \cite{Henkelman2000}, which is an improved version of the
traditional NEB method \cite{Jonsson1998} for finding minimum energy reaction
paths between two known minimum energy sites, resulting in a more accurate
estimation of the activation energy than the regular NEB method does, by
driving up one of the intermediate states near the transition point to reach
the highest saddle point along the reaction path.

For AIMD calculations, the Born-Oppenheimer approximation is used; ions are
considered as classical objects moving on the potential surface created by the
electrons obeying the quantum mechanical equation. The canonical ensemble
using the Nos\'{e} thermostat \cite{Nose1991} is employed. Electronic energies
are converged up to 10$^{-6}$ eV. With a time step of 1 fs, initial
equilibration steps are performed over 1 ps, while production runs are 6 ps long.

\section{Results and discussion}

\subsection{Determination of bulk properties of ceria}

\begin{figure}[ptb]
\begin{center}
\includegraphics[width=0.8\linewidth]{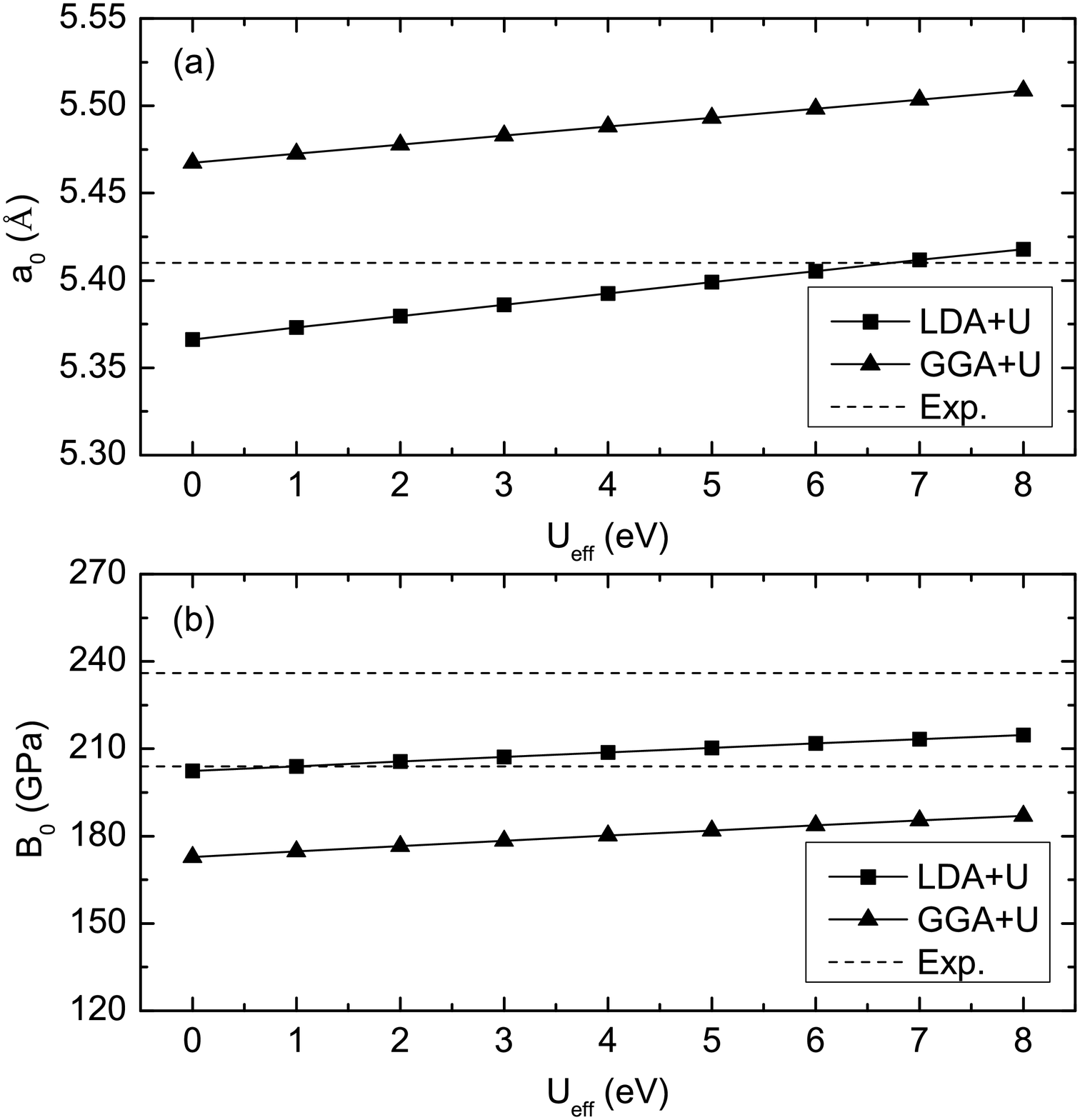}
\end{center}
\caption{Dependence of the lattice parameter (a) and bulk modulus
(b) of CeO$_2$
on U$_\text{eff}$. The dotted lines stand for the experimental measurements.}%
\label{fig1}%
\end{figure}

The stoichiometric ceria crystallizes in a CaF$_{2}$-like ionic
structure with space group F\emph{m}\={3}\emph{m}. The
experimentally determined lattice parameter is about 5.41
\r{A}\cite{Duclos1988,Gerward1993}, and the bulk modulus varies over
a rather broad range of 204 to 236 GPa
\cite{Duclos1988,Gerward1993,Gerward2005,Nakajima1994}. In this
section we demonstrate the dependence of the lattice parameter
$a_{0}$ and bulk modulus \emph{B} on U$_{\mathrm{eff}}$ in the range
$0\sim8$ eV, obtained by fitting the third-order Birch-Murnaghan
equation of state \cite{Birch1947}, as presented in Fig. \ref{fig1}.

We can see from Fig. \ref{fig1} that the reasonable lattice
parameter values $a_0$ are 5.40 \AA\ (LDA+\emph{U},
U$_{\mathrm{eff}}$=5 eV) and 5.48 \AA\ (GGA+\emph{U},
U$_{\mathrm{eff}}$=3 eV), and the bulk modulus values \emph{B} are
210.3 GPa (LDA+\emph{U}) and 178.4 GPa (GGA+\emph{U}). Noticeably,
our calculations are consistent with previous studies of Andersson
\emph{et al.} \cite{Andersson2007} and Loschen \emph{et al.}
\cite{Loschen2007}, but we can not reproduce the results obtained by
Jiang \emph{et al.} \cite{Jiang2005}. It is clear that the
GGA+\emph{U} calculations overestimate the lattice parameter and
underestimated the bulk modulus, and thus give a slightly less
accurate description of the atomic structure of ceria. On the other
hand, LDA+\emph{U} proposes a reasonable description. Moreover,
within the latter choice, the calculated O2\emph{p}-Ce4\emph{f} and
O2\emph{p}-Ce5\emph{d} band gaps are 2.21 and 5.29 eV, consistent
with the experimental measurements of about 3 and 6 eV, respectively
\cite{Wuilloud1984}. Moreover, it has been revealed that the LDA
treatment is actually better than the GGA ones when simulating
chemical reactions on reduced ceria surfaces \cite{Fabris2005}. In
addition, as mentioned above, most of the previous DFT calculations
dealt with the H$_{2}$O-ceria interactions by using GGA functional
\cite{Kumar2006,Chen2007,Fronzi2009,Yang2010}, with somewhat
conflicting results existing, hence the LDA treatment can be
performed as a beneficial complement. Therefore, in the following
surface calculations, we will present only the LDA+\emph{U} results.

\subsection{H$_{2}$O adsorption on stoichiometric ceria(111)}

The structural and energetic parameters of the free water molecule
are calculated within a box with the same size of the adsorbed
systems. The optimized geometry for free H$_{2}$O gives a bond
length of 0.97 \AA ~ and a bond angle of 105.2$^{\circ}$, consistent
with the experimental values of 0.96 \AA ~ and 104.4$^{\circ}$
\cite{Eisenberg1969}.

The adsorption energy of the system is calculated as follows:
\begin{equation}
E_{\mathrm{ad}}=E_{\mathrm{H_{2}O/ceria(111)}}-E_{\mathrm{H_{2}O}}%
-E_{\mathrm{ceria(111)}}, \label{Ead}%
\end{equation}
where $E_{\mathrm{H_{2}O}}$, $E_{\mathrm{ceria(111)}}$, and $E_{\mathrm{H_{2}%
O/ceria(111)}}$ are the total energies of the H$_{2}$O molecule, the
clean ceria(111) surface, and the adsorption system respectively.
According to this definition, a negative value of $E_{\mathrm{ad}}$
indicates that the adsorption is exothermic (stable) with respect to
a free H$_{2}$O molecule and a positive value indicates endothermic
(unstable) reaction.

\begin{figure}[ptb]
\begin{center}
\includegraphics[width=0.8\linewidth]{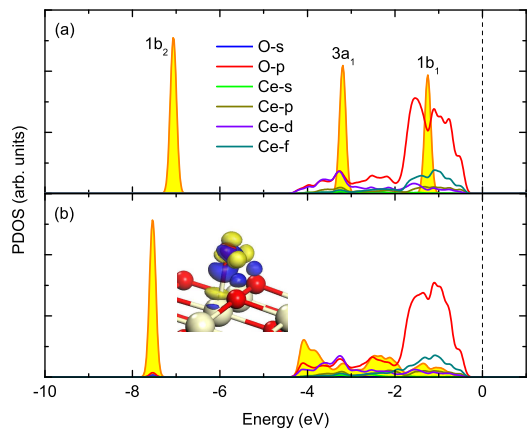}
\end{center}
\caption{(Color online) PDOS for the H$_{2}$O molecule and the
first-layer O and the second-layer Ce of the stoichiometric surface
for (a) free H$_{2}$O and the clean ceria(111) surface, (b) the A1
adsorption site. The inset in (b) shows the 3D electron density
difference, with the isosurface value set at $\pm$0.05
\textit{e}/\AA $^{3}$. The area filled with yellow color
represents MOs of H$_{2}$O. The Fermi level is set to zero.}%
\label{fig2}%
\end{figure}

Adsorption geometry optimizations are performed for a variety of
initial orientations of the H$_{2}$O molecule placed over different
surface sites, i.e., the flat-lying (water molecule lying parallel
to the surface), O-down (O atom pointing to the surface) and H-down
(two H atoms pointing to the surface) structures. We identify
different H$_{2}$O adsorption structures, whereas only the
low-energy one (A1) is discussed here, with the adsorption energy of
$-$0.96 eV and the corresponding frequencies of 3269.7, 3206.8, and
1490.3 cm$^{-1}$. The equilibrium geometry of A1 state is
illustrated in the inset of Fig. \ref{fig2}, in which the adsorbed
H$_{2}$O monomer lies nearly parallel to the surface, with the O
atom of water located almost on top of the second-layer Ce atom,
while H atoms symmetrically oriented pointing to the first-layer O
atoms ($d_{\mathrm{O-H}}$=1.85 \AA), namely, with two-H-bond
configuration formed, in agreement with the results of Fronzi
\emph{et al.} \cite{Fronzi2009}. Noteworthily, this adsorption
configuration is quite different from previous studies, either the
arbitrary assumption of the no-H-bond \cite{Henderson2003} or the
one-H-bond \cite{Kumar2006,Chen2007,Yang2010} configuration. The
difference may arise from that those previous studies underestimated
the hydrogen bond strength, which plays a key role in the
interactions between water and oxide surfaces. Moreover, the
dipole-dipole interactions between adsorbed water molecules at high
coverage probably blur the exact monomer adsorption configuration
studied in this letter.

In order to further understand the precise nature of the chemisorbed
molecular state, we plot in Fig. \ref{fig2} the electronic projected
density of states (PDOS) of the molecular adsorption structure for
the stable adsorption configuration of A1 in comparison with those
of free H$_{2}$O molecule and clean ceria(111) surface. The
three-dimensional (3D) electron density difference
$\Delta\rho(\mathbf{r})$, which is obtained by subtracting the
electron densities of noninteracting component systems, $\rho
^{\rm{ceria(111)}}(\mathbf{r})+\rho^{\rm{H_{2}O}}(\mathbf{r})$, from
the density $\rho(\mathbf{r})$ of the H$_{2}$O/ceria(111) system,
while retaining the atomic positions of the component systems at the
same location as in H$_{2}$O/cerai(111), is also shown in the inset
of Fig. \ref{fig2} (b). Positive (blue) $\Delta\rho(\mathbf{r})$
indicates accumulation of electron density upon binding, while a
negative (yellow) one corresponds to electron density depletion.
Molecular orbital (MO) 2$a_{1}$ of water (not shown here) is far
below the Fermi level and thus remains intact in water-metal
interaction. Here we consider only three MOs 1$b_{2}$, 3$a_{1}$, and
1$b_{1}$.

As illustrated in Fig. \ref{fig2}, upon adsorption, the orbital
1$b_{2}$ is observed to shift down in energy by 0.47 eV. More
significantly, the orbitals 3$a_{1}$ and 1$b_{1}$ undergo noticeable
changes, especially for the latter, becoming more broadened and
hence apparently delocalized, with the domain of energy
corresponding to that of the ceria surface. Obviously, the \emph{d}-
and \emph{f}-states of the Ce atom do not change much because of
localization. On the other hand, the shifting up in energy for the
peak of O-2\emph{p} states of the ceria surface play a key role in
the water-ceria interaction, which indicates that the adsorbed MO
1$b_{1}$ may act as an electron donor state. The features of the
orbital hybridization are further substantiated by the 3D electron
density difference plotted in the inset of Fig. \ref{fig2}(b). We
can see that there exists a large charge accumulation beneath the O
atom of the adsorbate. This accumulation should come from the charge
redistribution because of delocalization of the MOs of water, rather
than from the substrate, which is consistent with above PDOS
analysis demonstrating the localization of Ce electronic states.
Moreover, obvious charge transfer from H atoms of the adsorbate to
the surface O atoms is observed, suggesting strong hydrogen-bond
interactions between the adsorbate and substrate.

\begin{figure}[ptb]
\begin{center}
\includegraphics[width=0.8\linewidth]{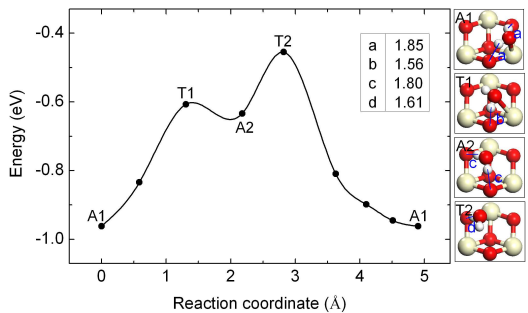}
\end{center}
\caption{(Color online) Variation in H$_{2}$O adsorption energy on
the stoichiometric ceria(111) surface as a function of the lateral
displacements of O atom from its original adsorption site A1 to the
nearest-neighbor one. Insets show the corresponding geometry
sketches for minima (A1, A2) and transition (T1, T2) states. The
chart summarizes the targeted O-H distances (in \AA) pointed out in
the insets. Large, moderate and small spheres stand
for Ce, O and H atoms, respectively.}%
\label{fig3}%
\end{figure}

Given a H$_{2}$O molecule at a stable adsorption site, it is
interesting to see how it diffuses on the substrate. Therefore, we
calculate the diffusion path and energetic barrier of water on
ceria(111) surface between neighboring stable A1 sites, with the
adsorption energy as a function of the lateral displacement of O
atom of the water molecule shown in Fig. \ref{fig3}. As indicated,
the molecular A1 state is adsorbed by $-$0.96 eV. Along the
diffusion path, the system evolves through an energy barrier of 0.35
eV to reach a semi-stable state A2, which is also nearly flat on the
surface. A2 is less stable with an adsorption energy of $-$0.63 eV,
with the O atom of water located on top of the third-layer O atom,
while as in the A1 state, two hydrogen bonds formed between H atoms
of the water and the surface-layer O atoms ($d_{\mathrm{O-H}}$=1.80
\AA). At the transition state T1, one H atom of the water points to
the top O atom, with the distance $d_{\mathrm{O-H}}$=1.56 \AA. Along
this path, the diffusion from A2 to the neighboring A1 is hindered
by a barrier of 0.18 eV, across the transition state T2, which also
forms single hydrogen bond. It can be seen from the transition
states T1 and T2 that the single hydrogen bond formed between the
water monomer and the oxide surface is unfavored in energy, which is
different from the previous high-coverage studies \cite{Kumar2006}
but consistent with our finding that the two-hydrogen-bond
interaction is favored for the monomer adsorption.

According to the overall energetic profile obtained above, we
calculate the diffusion coefficient at different values of
temperature \emph{T} of this system within the Arrhenius-type
relation, $D=\upsilon_{0}d^{2}e^{-\varepsilon/k_{B}T}$, where
$\upsilon_{0}$=3269.7 cm$^{-1}$ is the vibrational frequency of A1
state, $d$=4.90 \AA~is the displacement of the water during
diffusion, and $\varepsilon$=0.51 eV is the activation energy. For
comparison the diffusion coefficient at \emph{T}=300 K and 600 K are
calculated to be $7.18\times10^{-14}$ and $1.30\times10^{-9}$
m$^{2}$/s, respectively. It is clear that at 300 K, the coefficient
is too small to facilitate the diffusion process. While at 600 K,
diffusion should be observed easily.

\begin{figure}[ptb]
\begin{center}
\includegraphics[width=0.8\linewidth]{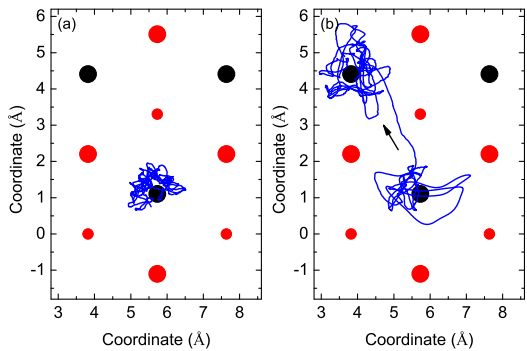}
\end{center}
\caption{(Color online) Projection on the stoichiometric ceria(111)
surface of the position of the oxygen atom (blue) corresponding to
the H$_{2}$O from AIMD simulation at the temperature of 300 K (a)
and 600 K (b), respectively. The black arrowhead in (b) denotes the
diffusion orientation. The large red and black filled circles mark
the equilibrium positions of the first-layer O atoms and the
second-layer Ce atoms, respectively, while the smaller red ones
represent the third-layer O atoms.}%
\label{fig4}%
\end{figure}

This is confirmed by our AIMD simulations, starting from the
adsorbed molecular site A1 at 300 K and 600 K, which are shown in
Fig. \ref{fig4}. At room temperature (300 K), it can be seen that
the molecular state A1 is recovered with only small oscillations.
Apparently the H atom oscillates more intensely than the O atom,
with the longest $d_{\mathrm{O-H}}$=1.54 \AA~in the water molecule.
At higher temperature of 600 K, the water molecule undergoes more
considerable oscillations (with the longest $d_{\mathrm{O-H}}$=1.63
\AA), and finally negotiates a diffusion to the neighboring
adsorption site A1, as identified in the above CI-NEB calculations.

Moreover, during AIMD simulations we find that at temperature as
high as 600 K, neither desorbed nor dissociated water configuration
is observed. This indicates that on one hand the water-oxide
interaction is strong enough to stand for high temperature, on the
other hand, the stoichiometric ceria surface is inactive towards
H$_{2}$O splitting. Actually, by static calculations, we find that
the co-adsorption of H and OH species on the regular ceria(111)
surface is endothermic thus unstable. In a word, the stoichiometric
ceria(111) surface does not dissociate water, in good agreement with
previous calculations \cite{Kumar2006,Chen2007,Fronzi2009,Yang2010}.
However, the existence of defects, especially the oxygen vacancies
on the surface of ceria, may act as active sites for the
dissociation of water. Hence, in the following section we will focus
our attention on this topic.

\subsection{H$_{2}$O adsorption on reduced ceria(111)}

\begin{figure}[ptb]
\begin{center}
\includegraphics[width=0.8\linewidth]{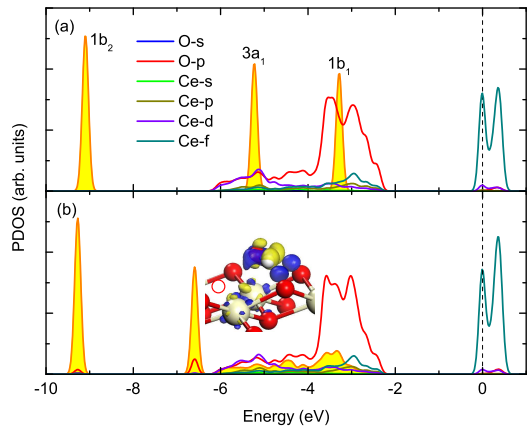}
\end{center}
\caption{(Color online) PDOS for (a) free H$_{2}$O and the clean
reduced ceria(111) surface, (b) the A2${'}$ adsorption site. The
isosurface value for the 3D electron density difference is set at
$\pm$0.05 \textit{e}/\AA $^{3}$. The red open circle in the inset
of (b) denotes the oxygen vacancy position. The Fermi level is set to zero.}%
\label{fig5}%
\end{figure}

It has been proposed that for bulk ceria, the oxygen-vacancy
formation process is facilitated by a simultaneous condensation of
two electrons into localized \emph{f}-level traps on two cerium
atoms \cite{Skorodumova2002}. The situation is similar for ceria
surface \cite{Yang2004}. We can see from Fig. \ref{fig5}(a) that in
the reduced surface the Ce-4\emph{f} states become partially
occupied, meanwhile other states of the surface remain almost
intact, compared with that in stoichiometric surface. This
observation may contribute much to the adsorption properties of
water on the reduced ceria surface.

\begin{figure}[ptb]
\begin{center}
\includegraphics[width=0.8\linewidth]{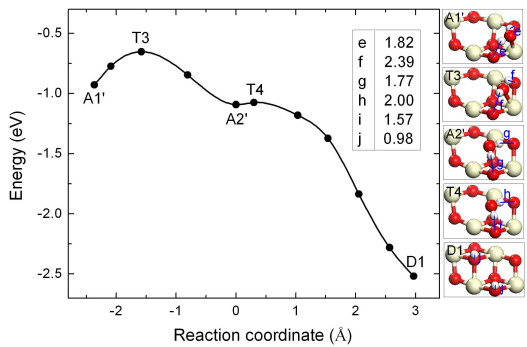}
\end{center}
\caption{(Color online) Variation in energy for the diffusion and
dissociation
of H$_{2}$O molecule on the reduced ceria(111) surface.}%
\label{fig6}%
\end{figure}

Similarly to the stoichiometric surface, we explore different
H$_{2}$O adsorption structures and find that the most stable
molecular adsorption state is A2 adjacent to the defect (labeled
A2${'}$) with the adsorption energy of $-$1.09 eV, while A1 adjacent
to the defect is the next most stable state (labeled A1${'}$) with
the adsorption energy of $-$0.93 eV. As illustrated in the insets of
Fig. \ref{fig6}, for these two stable molecular states, the
distances between the H atoms of water and first-layer O atoms are
1.77 and 1.82 \AA, respectively, both smaller than that on the
stoichiometric surface. In A2${'}$ configuration the O atom of the
adsorbed water is located on top of the third-layer O atom, with a
lower adsorption energy compared with that on the stoichiometric
surface. Apparently, as depicted by previous studies
\cite{Kumar2006,Chen2007,Fronzi2009,Yang2010}, the oxygen vacancy
enhances the water-ceria interaction.

Figure \ref{fig5}(b) presents the PDOS of the molecular state
A2${'}$. One can see that the MOs 1$b_{2}$ and 3$a_{1}$ keep
localized, only with a down-shift in energy by 0.16 and 1.36 eV,
respectively, while 1$b_{1}$ becomes totally delocalized, which
gives the main contribution to the adsorption behavior of water
monomer. In addition, the other contribution to the adsorption
behavior comes from the O-2\emph{p} states of the surface, and the
water adsorption introduces new peak for 2\emph{p} states of the
surface O atom, aligning in energy with 3$a_{1}$. From the 3D
electron density difference, we can observe more obvious charge
transfer from H atoms of the water to the surface O atoms, compared
with that on stoichiometric surface, probably providing a reason why
the water molecule adsorbs to the reduced surface more strongly.
Resembling that on the stoichiometric surface, the partially filled
Ce 4\emph{f} state almost does not participate the interaction with
water, which exactly reveals the importance of the Hubbard parameter
\emph{U} for correctly depicting its localization. Moreover, we
notice that this parameter prominently dominates the bulk properties
of ceria such as lattice constant, hence indirectly holds the
interactions between water and substrate, particularly for the
reduced ceria(111) surface.

To clarify how an oxygen vacancy affects the dynamical properties of
the adsorbed water monomer, let us examine the diffusion and
dissociation of the H$_{2}$O molecule on this reduced surface. We
calculate the diffusion path of the water between the stable states
A2${'}$ and A1${'}$ (see Fig. \ref{fig6}). It is clear that the
diffusion from A2${'}$ to A1${'}$ needs to overcome an energy
barrier of 0.44 eV, while the inverted diffusion from A1${'}$ to
A2${'}$ is more facilitated with a lower barrier of 0.27 eV.
Noticeably, the diffusion energy barriers presented here are larger
than that on the stoichiometric ceria surface, and this should be
attributed to the rotation of H$_{2}$O molecule during the diffusion
process, which can be seen from the transition state T3 that has an
O-down structure.

\begin{figure}[ptb]
\begin{center}
\includegraphics[width=0.8\linewidth]{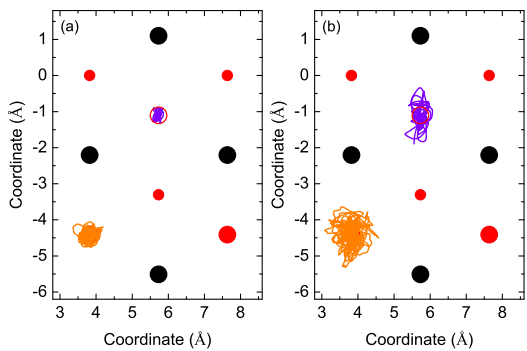}
\end{center}
\caption{(Color online) Projection on the reduced ceria(111) surface
of the position of the oxygen atom corresponding to the OH species
(purple) and the hydrogen atom (orange) from AIMD simulation at the
temperature of 100 K (a) and 600 K (b), respectively. The red
open circle denotes the oxygen vacancy position.}%
\label{fig7}%
\end{figure}

To determine the final configurations of the dissociation, we
performed AIMD simulations for water on the reduced ceria(111)
surface, starting from the adsorbed molecular site A2${'}$, at
different temperatures (see Fig. \ref{fig7}). What is interesting,
even at temperature as low as 100 K, is that the water monomer first
oscillates around its equilibrium position and soon starts to
dissociate into H and OH species almost \emph{in situ}. It is found
that the O atom of the OH species occupies the position of oxygen
vacancy, with the O--H bond almost perpendicular to the surface,
oscillating together with the surface. Whereas, another oscillating
H atom of the dissociated water molecule is bonded to the adjacent
top-layer O atom. Namely, hydroxyl surface forms upon the
dissociation of water on the reduced ceria surface. This is the same
at higher temperature of 600 K, only with a more intensive
oscillation.

With this final dissociation configuration, now we return to the
static CI-NEB calculations. As presented in Fig. \ref{fig6}, the
adsorption energy of the dissociated water state D1 was calculated
to be $-$2.52 eV, with the bond length between H and O of 0.98 \AA.
To reach the state D1, the system only needs to overcome a small
barrier, 0.02 eV, corresponding to the low temperature 100K, as
depicted in the AIMD simulations. At the transition state T4 leading
to the dissociation, the interatomic distance between the O atom of
the OH species and another H atom is quite small, 1.12 \AA. These
results consist well with previous experimental observation
\cite{Kundakovic2000} and theoretical calculation \cite{Yang2010},
while differ from the calculation with plain DFT \cite{Fronzi2009}
that resulted in a high activation energy 2.35 eV for water
dissociation.

Interestingly, by static calculations we have also identified a
semi-stable adsorption configuration with the adsorption energy
$-$0.89 eV, in which the O atom of the water occupies the position
of the oxygen vacancy, with a no-H-bond configuration formed. This
structure has once been considered as the most stable one on the
reduced ceria surface \cite{Yang2010,Fronzi2009}, and also a
specious initial state for evolving hydrogen gas \cite{Kumar2006} at
high coverage (0.5--1.0 ML). However, we find that the process of
H$_{2}$ molecule production is energetically unfavorable when
compared with the formation of hydroxyl surface (about 0.52 eV
higher in energy), therefore it cannot be observed in the AIMD
simulations. Furthermore, it has been pointed out that the energy
needed for H$_{2}$ production is as high as 2.05 eV \cite{Yang2010},
consistent with our results and the experimental observation that
the reduced ceria(111) surface is resistant to being oxidized by
water \cite{Henderson2003}.

\section{Conclusions}

In conclusion, we have systematically studied the adsorption and
dissociation behaviors of H$_{2}$O monomer on ceria(111) surface by
first-principles calculations complemented with AIMD simulations.
The most stable adsorption configurations on both the stoichiometric
and reduced surfaces were identified, with the electronic structures
analyzed in detail. Blocked by a diffusion barrier of 0.51 eV, water
on the stoichiometric surface is very immobile up to at room
temperature, while at higher temperature of 600 K, the water
molecule can freely diffuse on the surface. For the reduced surface,
We have found that the reaction of water and ceria is enhanced by
the on-surface oxygen vacancy. It is facilitated for water to
dissociate into H and OH species, only hindered by a small barrier
of 0.02 eV, which was confirmed by the molecular dynamics
simulations at temperature as low as 100 K. Besides, the reduced
ceria(111) surface is resistant to being oxidized by water, in good
accordance with the existing experimental results. The present
results obtained at low water coverage afford to settle existing
discrepancies about the behaviors of water on ceria surface, and
provide a guiding line for deeply understanding the water-ceria
surface interactions.

\begin{acknowledgments}
This work was supported by NSFC under Grants No. 51071032 and No.
60821061.
\end{acknowledgments}

\end{document}